\documentclass[prd,twocolumn,floatfix,preprintnumbers,showpacs]{revtex4}
\usepackage{graphicx}% Include figure files
\usepackage{dcolumn}% Align table columns on decimal point
\usepackage{bm}% bold math
\usepackage{color}
\usepackage{hyperref}
\usepackage{amsmath} % Using 'align' in stead of 'eqnarray'.
\def\lsim{\mathrel{\mathop
  {\hbox{\lower0.5ex\hbox{$\sim$}\kern-0.8em\lower-0.7ex\hbox{$<$}}}}}
\def\gsim{\mathrel{\mathop
  {\hbox{\lower0.5ex\hbox{$\sim$}\kern-0.8em\lower-0.7ex\hbox{$>$}}}}}

\def\aj{AJ}
\def\apj{ApJ}
\def\aap{A\&A}
\def\apjl{ApJ}

\def\PhysRev{PhysRev}

\def\mnras{MNRAS}

\def\pasj{PASJ}
\def\prd{PRD}

\newcommand{\nc}{\newcommand}

\nc{\be}[1]{\begin{equation}\mbox{$\label{#1}$}}
\nc{\bea}[1]{\begin{eqnarray}\mbox{$\label{#1}$}}

\nc{\Section}[2]{\section{#2}\label{#1}}
\nc{\Bibitem}[1]{\bibitem{#1}}
\nc{\Label}[1]{\label{#1}}

\nc{\Mpc}{Mpc/h}
\nc{\vev}[1]{\langle #1 \rangle}

\nc{\eea}{\end{eqnarray}}
\nc{\ee}{\end{equation}}

\begin{document}

\preprint{astro-ph/...}

\title{Clustering of Luminous Red Galaxies III:\\
Detection of the Baryon Acoustic Peak in the 
3-point Correlation Function}

\author{Enrique Gazta\~naga}
\email{gazta@ice.cat}
\affiliation{Institut de Ci\`encies de l'Espai, IEEC-CSIC, 
F. de Ci\`encies, Torre C5 par-2,  Barcelona 08193, Spain}

\author{Anna Cabr\'e}
\affiliation{Institut de Ci\`encies de l'Espai, IEEC-CSIC, 
F. de Ci\`encies, Torre C5 par-2,  Barcelona 08193, Spain}
  
\author{Francisco Castander}
\affiliation{Institut de Ci\`encies de l'Espai, IEEC-CSIC, 
F. de Ci\`encies, Torre C5 par-2,  Barcelona 08193, Spain}
 
\author{Martin Crocce}
\affiliation{Institut de Ci\`encies de l'Espai, IEEC-CSIC, 
F. de Ci\`encies, Torre C5 par-2,  Barcelona 08193, Spain}

\author{Pablo Fosalba}
\affiliation{Institut de Ci\`encies de l'Espai, IEEC-CSIC, 
F. de Ci\`encies, Torre C5 par-2,  Barcelona 08193, Spain}

\date{\today}

\pacs{98.80.Cq}
\begin{abstract} 
We present the 3-point function $\xi_3$ and  $Q_3=\xi_3/\xi_2^2$
for a spectroscopic sample of
luminous red galaxies (LRG) from the Sloan Digital Sky Survey DR6 \& DR7.
We find a strong (S/N$>$6) detection of $Q_3$ on scales of 
55-125 Mpc/h, with a well defined peak around 105 Mpc/h
in all $\xi_2$, $\xi_3$ and $Q_3$,
in excellent agreement with the predicted shape and location of the 
imprint of the baryon acoustic oscillations (BAO). We use
very large simulations (from a cubic box of L=7680 Mpc/h) to asses and test the
significance of our measurement.
Models without the BAO peak are ruled out by the $Q_3$
data with 99\% confidence.
This detection demonstrates the non-linear growth of structure  by gravitational 
instability between $z=1000$ and the present. Our measurements  show the expected
shape for $Q_3$ as a function of the triangular configuration. This
provides a first direct measurement  of  the non-linear mode coupling coefficients 
of density and velocity fluctuations which, on these large scales,
are independent of cosmic time, the amplitude 
of fluctuations or cosmological parameters. 
The location of the BAO peak in the data indicates
$\Omega_m =0.28 \pm 0.05$ and $\Omega_B=0.079 \pm 0.025$ (for
$h=0.70$) after marginalization over  spectral index ($n_s=0.8-1.2$)
linear $b_1$ and quadratic $c_2$ bias,
which are found to be in the range: $b_1=1.7-2.2$
and $c_2=0.75-3.55$.  The data  allows a hierarchical contribution from
primordial non-Gaussianities in the range $Q_3=0.55-3.35$.
These constraints are independent and complementary
to the ones that  can be obtained using the 2-point function, which are presented in a
separate paper. This is the first detection of the shape of $Q_3$ on BAO 
scales, but our errors are shot-noise dominated and the SDSS volume
is still relatively small, so there is ample room for future 
improvement in this type of  measurements.
\end{abstract}

\maketitle

\section{Introduction}

The galaxy  three-point function $\xi_3$ provides a crucial test and
 valuable statistical tool  to investigate the origin of structure formation
and the relationship between galaxies and
 dark matter (see  \citet{bernar} for a review). We will concentrate
here on the reduced 3-point function $Q_3 \simeq \xi_3/\xi_2^2$ defined in Eq.\ref{fiftheq}
as the scaling expected from non-linear couplings (with $Q_3 \simeq 1$).
Measurements of the three-point function and other higher-order 
statistics in galaxy catalogs have a rich history (eg \citet{peeblesgroth}, 
\citet{frypeebles}, \citet{baumgartfry}, \citet{gazta1992}, 
\citet{bouchet1993}, \citet{frygazta1994}). 
In the past decade, three-point statistics have confirmed the basic 
picture of gravitational instability from Gaussian initial conditions  
(\citet{friemangazta1994}, \citet{jingborner1998}, \citet{friemangazta1999},  
\citet{feldman2001}). The connection between these observables and theoretical
predictions is best  done on large scales, where the physics (gravity)
is best understood, 
but the surveys  previously available were not large enough to have good statistics
on sufficiently large scales.

With the completion of large redshift surveys such as 2dFGRS 
(\citet{colless2001}) and SDSS (\citet{York00}) we expect
 measurement of higher-order  statistics 
to provide tighter constraints on cosmology
(\citet{colombietal1998}, \citet{szapudietal1999}, \citet{matarreseetal1997}, 
\citet{scoccimarro2004}, \citet{sefusatti2005}).
First measurements of the redshift space $\xi_3$ in the 
2dFGRS (\citet{gazta2005}) and SDSS (eg \citet{nichol2006})
show good agreement with expectations
(see also \citet{kayo2004}, \citet{nishimichi2007}, \citet{kulkarni2007}
and references therein).

We will assume here that the initial conditions are Gaussian. Current models of 
structure formation predict a small level of initial non-Gaussianities that
we can neglect here. A popular 
parametrization of this effect is to assume that initial curvature perturbations,
given by the  gravitational potential, $\Phi$,  are given by 
$\Phi = \Phi_L + f_{NL} ~(\Phi_L^2 - <\Phi_L^2>)$, where $\Phi_L$ is
Gaussian and $f_{NL}$ is a non-linear coupling parameter of order unity. This
produces non-Gaussianities in the matter density perturbations at wavenumber
$k$ which, using the Poisson equation, are suppressed by the square of the horizon scale 
$k_H \equiv H_0/c$, so that:  $Q_3 \simeq 3 f_{NL} (k_H/k)^2 T(k)/D(a) $, where $T(k) \simeq 1$
is the so called CDM transfer function and $D(a)$ is the growth factor. 
In our analysis $(k_H/k)^2 \simeq 10^{-3}$ so that these type 
of primordial non-Gaussianities produce negligible contribution to 
$Q_3$ for models with $f_{NL} \simeq 1$. 
A more detailed analysis of this will be presented
elsewhere (see \citet{sefusattikomatsu} 
for a detailed forecast for this model). If the non-Gaussianities
come from a non-linear coupling in the matter density field rather than in
the gravitational potential, the resulting 3-point function will have
a non-Gaussian contribution similar to that produced by non-linear bias
$c_2$ (ie see Eq.\ref{eq:Q3G} below  and Conclusions).

The shape and amplitude of $Q_3$ depends on galaxy bias, 
ie how galaxy light traces the dark matter (DM) distribution. 
This is both a problem and an opportunity. A problem because
biasing can confuse our interpretation of the observations.
An opportunity because one can try to measure the biasing parameters
out of $Q_3$. This idea was first proposed by \citet{frygazta1993}
and has been applied to the 3-point function and bispectrum of  real data 
(\cite{friemangazta1994,fry1994,feldman2001,scoccimarro2001,verdeetal2002,gazta2005,nichol2006}).

This paper is the third on a series of papers on clustering of LRG. In the
first two papers \cite{paper1,paper2} we studied redshift space distortions on the 2-point 
correlation function. The reader is referred to these papers for more details
on the LRG samples, simulations and the systematic effects. Similar LRG samples 
from SDSS have already been used by different groups to study the 2-point function
(eg \cite{detection,hutsia,percival,padmanabhan2007,blake}) and found good agreement
with predictions in the BAO scales, where density fluctuations are at a level
of few percent. This is encouraging and indicates that this sample is large and accurate
enough to investigate clustering on such large scales.

In this paper we follow closely the methodology presented in 3 previous analysis.
\citet{barrigagazta2002}  presented a comparison of the predictions for the two and three-point 
correlation functions of density fluctuations, $\xi_2$ and $\xi_3$, in
gravitational perturbation theory (PT) against large Cold Dark Matter (CDM) simulations.
Here we use the same method and codes to estimate the clustering in simulations.
 \citet{gaztascocci2005} 
extend these results into the non-linear regime and focus on the effects
of redshift distortions and the extraction of galaxy bias parameters in galaxy surveys. 
\citet{gazta2005} apply this methodology to the 2dFGRS. Here we apply the very same
 techniques to the LRG data, so the reader is referred to these papers for more
details.

\citet{kulkarni2007} have also estimated $\xi_3$
 using LRG galaxies from SDSS DR3, but focusing on smaller scales.
We use DR6  which has 3 times the area (and volume) of DR3. We also use 
a volume limited sample and a different estimator for the correlation
functions and errors, focusing on the largest scales.

\section{Theory}

\subsection{Definitions}

The two and three-point correlation functions are defined, respectively, as
\bea{xi2xi3}
\xi_2(r_{12}) &=& \langle \delta(r_1) \delta(r_2) \rangle \\
\xi_3(r_{12},r_{23},r_{13}) &=& \langle \delta(r_1) \delta(r_2) \delta(r_3) \rangle 
\eea
where $\delta(r) = \rho(r)/\bar{\rho}-1$ is the local density fluctuation about 
the mean $\bar{\rho}=\langle\rho\rangle$, and the expectation value is taken over different
realizations of the model or physical process. In practice, the expectation
value is over different spatial regions in our Universe, which are
assumed to be a fair sample of possible realizations (see Peebles 1980).
A possible complication with this approach is the so call finite volume
effects which result in potential estimation and ratio biases (eg see \cite{HuiGazta,bernar}).
For our samples we have checked using a large simulation
that these potential estimation biases are small compared to the errors.
This can be seen in Fig.\ref{fig:q3sim} below which shows a good agreement between
predictions and mock simulations that have the same size than the SDSS sample that we
used in our analysis.
We use a simulation with about 512  more volume than 
the data. We split this large simulation into 512 subsamples and estimate
the mean and error (from the variance) of the 2 and 3-point correlation 
functions in the 512 subsamples. We find that this mean agrees well, well within 
the error, with the corresponding correlation estimated in the full 
simulation. This indicates that the finite volume effects are negligible compared to errors
(see also Fig.5 in paper IV, \cite{paper4}).

It is convenient to define a  $Q_3$ parameter as \citet{grothpeebles1977}

\begin{eqnarray}
Q_3 &=& \frac{\xi_3(r_{12},r_{23},r_{13})}{\xi_3^H(r_{12},r_{23},r_{13})} \label{fiftheq}
\\
\xi_3^H  &\equiv& 
{\xi_2 (r_{12})\xi_2(r_{23})+\xi_2(r_{12})\xi_2(r_{13})+\xi_2(r_{23})\xi_2(r_{13})},
\nonumber
\end{eqnarray}

\noindent

where we have introduced a definition for the "hierarchical" three-point function $\xi_3^H$.
Note that, by homogeneity, the 3-point function $\xi_3$ or $Q_3$ can only depend on the
distance between $r_1$, $r_2$ and $r_3$. This involves 3 variables
that define the triangle formed by the 3 points. In principle $Q_3$ could
depend on the geometry and scale of the triangle.
Here we will use two of the triangle sides $r_{12}$, $r_{23}$ and 
$\mu$, the co-sinus of the angle between $\vec r_{12}$ and $\vec r_{23}$, 
which we call $\alpha$. Thus $Q_3=Q_3(r_{12}, r_{13},\mu)$.

The $Q_3$ parameter was thought to be roughly constant 
as a function of triangle shape and scale (\citet{peebles1980}), a result that is usually 
referred to as the hierarchical scaling.  Accurate measurements/predictions 
show that $Q_3$ is not quite constant in any regime of
clustering, although the variations of $Q_3$ with scale and shape are small compared to the
corresponding  changes in $\xi_2$ or $\xi_3$, specially at small scales.

\subsection{$Q_3$ and mode coupling}

We will illustrate next how measurements of $Q_3$
provide a direct estimation of  the non-linear mode coupling  
of density and velocity fluctuations.
First consider the fully non-linear fluid equations 
that determine the gravitational evolution of
density fluctuations, $\delta$,  and the divergence of the
velocity field, $\theta$, in an expanding universe
for a pressureless irrotational fluid. 
In Fourier space (see Eq.37-38 in \cite{bernar}):
\begin{eqnarray}
&\dot{\delta}+\theta = -\int dk_1 dk_1 \alpha(k_1,k_2)\theta(k_1)\delta(k_2)
 \label{eq:fluid}&
\\
&\dot{\theta}+H\theta+{3\over{2}} \Omega_m H^2\delta 
= -\int dk_1 dk_1 \beta(k_1,k_2)\theta(k_1)\theta(k_2) &
\nonumber
\end{eqnarray}
where derivatives are over time $d\tau=a dt$ and $H= d\ln a/d\tau$.
On the left hand side $\delta=\delta(k)$ and $\theta=\theta(k)$
are functions of the Fourier wave vector $k$.
The integrals are over vectors $k_1$ and $k_2$ constrained to $k=k_{12} \equiv k_2-k_1$.
The right hand side of the equation include the non-linear terms which are
quadratic in the field and contain the mode coupling functions:
\begin{equation}
\alpha= {k_{12}*k_1\over{k_1^2}}
~~ ~~;~~ ~~ 
\beta = {k_{12}^2 (k_2*k_1)\over{2 k_1^2 k_2^2}} 
 \label{eq:mode}
\end{equation}
where ``$*$'' is the scalar product of the vectors (Eq.39 in \cite{bernar}). 
This functions, $\alpha$ and $\beta$, account for the mixing of Fourier modes.
Note that this functions are adimensional and depend on the
geometry of the triangle formed by the two wave vectors $k_1$ and $k_2$ that contribute to
$k=k_2-k_1$. The scalar product indicates that 
mode couplings are larger when the modes are aligned. Physically
this means that density and velocity gradients created by (non-linear) 
gravitational growth will tend to be parallel.

Consider now the following perturbation expansions:
\begin{equation}
\delta = \sum_i ~\delta_i  ~~  ~~;~~ ~~ \theta = \sum_i ~\theta_i
\label{eq:series}
\end{equation}
 The first terms,  $\delta_1$ and $\theta_1$ in the above series are the 
 linear solution of the fluid equations Eq.\ref{eq:fluid},
 where we neglect the quadratic terms in the equations. The linear term
yields $\dot{\delta}_1=-\theta_1$ for the first fluid equation in Eq.\ref{eq:fluid}.
Combined  with the second fluid equation yields the well known harmonic oscillator
equation for the linear growth:
\begin{equation}
\ddot{\delta}_1+H\dot{\delta}_1-{3\over{2}} \Omega_m H^2\delta_1 = 0
\label{eq:harmonic}
\end{equation}
This equation is valid in Fourier or in configuration space.
In linear theory
each Fourier mode evolves independently of the other and they all grow
linearly out of the initial fields with the same growth function,
ie $\delta_1= D(t) \delta_0$, where $\delta_0$ is the value of the
field at some initial time 
and $D(t)$ is the linear growth function, which is a solution
to the above harmonic equation.
If the initial field $\delta_0$
is Gaussian then $\delta_1$ is also Gaussian. As shown
above mode coupling is a non-linear effect.  By construction, 
the next terms in the series of Eq.\ref{eq:series},
$\delta_2$ and $\theta_2$, are assumed to be quadratic in the 
linear terms, ie $\delta_2 \propto \delta_1^2$. The solution for the 
second order can be found by just replacing the above expansion into
the fluid equations  keeping the second order terms in the equation.
First order terms in the left side of the fluid equations in
Eq.\ref{eq:fluid}
vanish by construction of the linear equation.  We then
find (see Eq.156 in \cite{bernar}):
\begin{eqnarray}
\delta_2(k) &=& \int dk_1 dk_2 F_2(k_1,k_2) \delta_1(k_1)\delta_1(k_2)
 \label{eq:F2}
\\
F_2 &=& {5\over 7}\alpha+{2\over 7}\beta
\nonumber
\end{eqnarray}
Thus the second order term 
in the expansion is $\delta_2 \propto F_2 \delta_1^2$
and contains the mode coupling information through $F_2$.

Let's consider now the observables, which are the
2 and 3-point correlation functions. 
Note that for an initially Gaussian field 
$\delta_1$ is also Gaussian so that  $<\delta_1^3>=0$
and  $<\delta_1^4>= 3 <\delta_1^2>^2$. 
At leading order
the 2-point function $\xi_2$ is dominated by the linear
evolution and the second to leading order
term is zero because $<\delta_1 \delta_2>=F_2<\delta_1^3>=0$.
Thus we have $\xi_2(t) = D(t)^2 \xi_2(0)$.
The linear contribution to  $\xi_3$ is also
zero and the leading order in $\xi_3$ comes
from having $\delta_2$ in one of the 
3 points and $\delta_1$ in the other 2 points. 
For illustration, let us ignore for now the
arguments of the 3 points, and see how things
scale:
\begin{equation}
\xi_3 = <\delta_1 \delta_1 \delta_2> \propto F_2 <\delta_1^4>
 = 3 F_2 <\delta_1^2>^2 \propto F_2 \xi_2^2
\end{equation}
This is in agreement with the scaling $\xi_3 = Q_3 \xi_2^2$ in
Eq.\ref{fiftheq}.
We therefore have that $Q_3$ at leading order is just proportional
to $F_2$. This is exactly the case when we account for the
arguments in the 3 points, as can be easily checked.
The dependence of $F_2$ on the triangular configuration, ie
through Eq.\ref{eq:mode}, is all contained in the corresponding triangular 
configurations of $Q_3$, ie $Q_3=Q_3(r_{12}, r_{13},\mu)$.
This shows the interest of measuring $Q_3$: it
provides a direct way to measure the mode coupling information in $F_2$,
which includes the fully non-linear coupling in Eq.\ref{eq:mode}. Higher order
correlations just provide information about different combinations of
coupling functions $\alpha$ and $\beta$.

\subsection{Gravitational instability}

We can now address the following question: is large
scale structure produced by gravitational growth from
small Gaussian fluctuations? We could answer this
question by measuring $\xi_2(t)$. If we knew
some initial conditions $\xi_2(0)$, we can then
estimate $D(t)^2 \simeq \xi_2(t)/\xi_2(0)$ and
compare to the linear solution of Eq.\ref{eq:harmonic}.
By measuring $Q_3$ we can directly compare to $F_2$.
This is independent of the linear test in $D(t)$ and does not
require knowledge of the initial conditions $\xi_2(0)$.
The result should in fact be independent of time.

We can also test gravitational growth with independence of time
or initial conditions by
using the linear relation between density and
velocities,  $\dot{\delta}_1=-\theta_1$.
 But this is a test  of linear evolution and  requires measurements 
of the velocity field (this, in fact, is tested using redshift space distortions in Paper-I of this
series).  $Q_3$ only needs density fields and explores the non-linear
sector of gravity.

If the origin of structure is non gravitational (or gravity is
non-standard) but the initial
conditions are Gaussian,  then on dimensional grounds 
we would also expect 
$\xi_2(t) = D(t)^2 \xi_2(0)$ and $Q_3$ to be indepent of time.
But both the amplitude and shape of $Q_3$ as a function of
triangle configuration could be quite different if the fluid
equations are different, either because of a non-standard 
cosmology or non-standard law of gravity (eg see \cite{lobo,bernar}).
In the standard case, the shape of  $F_2$ is  mostly independent 
of cosmological parameters
(eg $\Omega_m$, $\Omega_\Lambda$ or $\Omega_b$),
time or the amplitude of fluctuations \cite{bernar}. 
As shown above, density and velocity gradients produced by
non-linear evolution are parallel  
which results in enhancement of clustering for elongated triangles.
The exact amplitude as a function of triangular shape
provides a finger print for non-linear gravitational growth.
It is purely a non-linear effect that is not
present in the initial conditions (which are assumed 
to be Gaussian with $Q_3=0$).  Unfortunately, the prediction
for $Q_3$ depends not only on the mode coupling $F_2$ but also
on the slope of the initial  $\xi_2(0)$. Models with relatively
more large scale power (ie smaller slope) produce structures with larger coherence 
which give rise to more anisotropic structures and stronger
shape dependence in $Q_3$. Fortunately the shape of $\xi_2$ can
also be estimated from data and one can then test
the mode coupling predictions for $F_2$, as we will show below.

\subsection{Shape dependence and BAO}

As mentioned above, there is a degeneracy in the shape of 
$Q_3$ between dark matter density
$\Omega_m$ and baryon density $\Omega_B$ because they produce degenerate slopes in $\xi_2$.
This degeneracy is broken by the presence of the BAO peak, but this requires a
measurement at 100 Mpc/h scales. To illustrate this, 
consider a  power law spectrum  $P(k)=Ak^n$. In this case
 (see  ~\cite{bernar} and  references therein):

\begin{eqnarray} 
\lefteqn{\xi_3(r_{12},r_{23},\mu )=
\left [ \frac{10}{7}+\frac{n+3}{n}\mu(\frac{r_{12}}{r_{23}}+\frac{r_{23
}}{r_{12}})+\right. } \\ 
& & \left.+\frac{4}{7}\frac{(3-2(n+3)+(n+3)^2\mu ^2)}{n^2}\right ]
 \xi_2(r_{12})\xi(r_{23})+\cal{P}\nonumber 
\label{sixtheq}
\end{eqnarray} 
where $\cal{P}$ stands for permutations of the indexes 123, and
 $\mu$  is the co-sinus of the angle between $\vec r_{12}$ and $\vec r_{23}$, 
which we call $\alpha$. Here we will only  show results as a function of
$\alpha$ for fixed $r_{12}$ and $r_{23}$. The above formula 
is only valid for  a power-law spectrum. For CDM we will use a full calculation
as explained below.

Elongated or ``collapsed configurations" are those with $\alpha \simeq 0$ or
$\alpha \simeq 180$ deg.
We  use the term "strong configuration dependence"  when there is a significant
difference between the collapsed and the perpendicular configurations. By 
"weak configuration dependence" we mean that $Q_3$ is ``hierarchical''
 (ie constant as a function of $\alpha$).
$Q_3(\alpha)$ flattens with the decrease of the spectral index $n$ and
for larger $n$ has a strong configuration dependence with a 
characteristic ``V'' shape. This corresponds to a  larger probability
of finding 3 points aligned, a direct consequence of gravitational infall
that enhances the filaments that are characteristic of large scale structure
both in simulations and real data.  In a CDM spectrum this dependence on $n$
translates into flat $Q_3$ (rounder structures) at small scales progressively getting
stronger V-shape as the effective $n$ gets larger because of the CDM transfer
function. At a fixed scale, the V-shape gets more pronounced when
we increase $\Omega_m$ or when we decrease $\Omega_B$. This is just
due to the effect of $\Omega_m$ and $\Omega_B$ in the CDM transfer function,
which changes the effective spectral index $n$. If we fix the shape
of the $P(k)$ spectrum, the values of $\Omega_m$ and $\Omega_B$ have very 
little effect on $Q_3$, as mentioned above.

\begin{figure}
\includegraphics[width=7.cm]{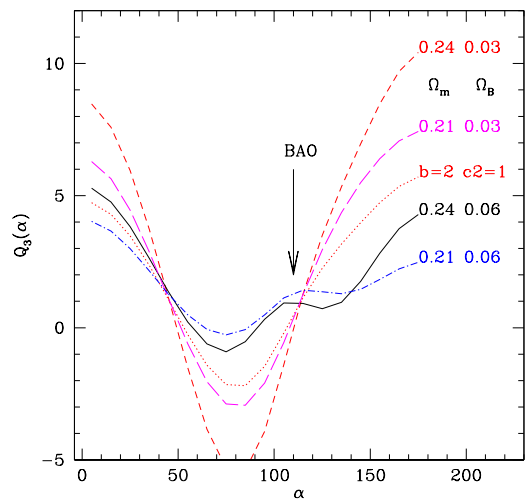}
\caption{Perturbation theory predictions for the
reduced 3-point function $Q_3$ for different values of $\Omega_m$
and $\Omega_B$, as labeled in the Figure. All cases are unbiased,
except for dotted line which corresponds
to $\Omega_m=0.24$ and $\Omega_b=0.03$ with $b_1=2$ and $c_2=1$.
This is for
triangles with two sides fixed to  $r_{12}=33\pm 5$ Mpc/h and  $r_{13}=88 \pm 5$ Mpc/h
as a function of the interior angle $\alpha$  between this two sides.
As $\alpha$ varies from $0-180$ degrees the third side of the
triangle changes from $r_{23}=55$ Mpc/h to  $r_{23}=121$ Mpc/h.
For high values of $\Omega_B$ a BAO peak (marked by the
arrow) shows at $\alpha=115$ degrees, which corresponds to 
 $r_{23}=106$ Mpc/h. The precise location depends on the cosmology used.}
\label{fig:q3th}
\end{figure}

These points are illustrated in Fig.\ref{fig:q3th}
which shows perturbation theory predictions for $Q_3$. For this
calculation we use Eq.(8) in \citet{barrigagazta2002}, where different scales
and triangular configurations are also shown. Here we focus on
the results of $Q_3$ around the BAO scales. For high values of
$\Omega_B$ a baryonic peak emerges in $Q_3$ and this peak
could be used as a cosmological prove or to break the $\Omega_m-\Omega_B$
degeneracy. The peak is present in all $\xi_2$, $\xi_3$ and $Q_3$.
As far as we know this is the first calculation illustrating 
how the BAO peak shows in $Q_3(\alpha)$.

Predictions in redshift space are harder to make (see \citet{gaztascocci2005}), 
but on large scales simulations show that redshift distortions in
$\xi_3$ and $\xi_2^2$ cancel out in $Q_3$, and we do not expect deviations
from PT tree-level predictions (see Fig.\ref{fig:q3sim} below).

\subsection{Biasing}

The value and shape of $Q_3(\alpha)$ changes with galaxy bias.
On very large scales we expect that 
galaxy fluctuations $\delta_G$ can be modeled 
as a local (but non-linear) function of the corresponding matter 
fluctuations $\delta$, so that $\delta_G \simeq F[\delta] $.
 For small fluctuations, $\delta\ll 1$, we can expand this local function as:

\begin{equation}
\delta_G \simeq F[\delta] \simeq \sum_{i}~{b_i\over{i!}}~\delta^i ,
\label{eq:bk}
\end{equation}
where $i=0$ comes from the requirement that $\langle \delta_G \rangle =0$. It then
follows (see \citet{frygazta1993}, \citet{friemangazta1994}) that:
\begin{eqnarray} 
\xi_2^G(r) & \simeq & b_1^2~\xi_2(r) \\
Q_3^G &\simeq& {1\over{b_1}} ~\left(Q_3+c_2 \right)
\label{eq:Q3G}
\end{eqnarray} 
where $c_2 \equiv b_2/b_1$, and the $\simeq$ sign indicates that this
is the leading order contribution in the expansion given by Eq.~(\ref{eq:bk}) above.  
Thus, in general, the linear bias prescription is not accurate for 
higher-order moments even when $\delta \ll 1$, the reason being 
that non-linearities generate non-Gaussianities of the same order as 
those of gravitational origin. The linear bias term $b_1$ can produce 
distortions in the shape of $Q_3$, while the non-linear terms $c_2$ 
only shifts the curve. This is illustrated by the dotted line in
Fig.\ref{fig:q3th}.
It is therefore possible, but challenging,  to use the shape of 
$Q_3^G$ in observations, when compared to the DM predictions, 
to separate $b_1$ from $c_2$ in the above relation. 
This gives an estimate of the linear bias $b_1$ which is independent of 
the overall amplitude of clustering (eg. $\sigma_8$).  This approach has already 
been implemented for the skewness $S_3$ \cite{gazta1994,gaztafrieman1994}, 
the bispectrum \cite{friemangazta1994,fry1994,feldman2001,verdeetal2002} 
or the angular 3-point function 
\cite{friemangazta1999} and has been used to forecast future analysis
\cite{sefusatti2006}.

\begin{figure*}
\includegraphics[width=7.cm]{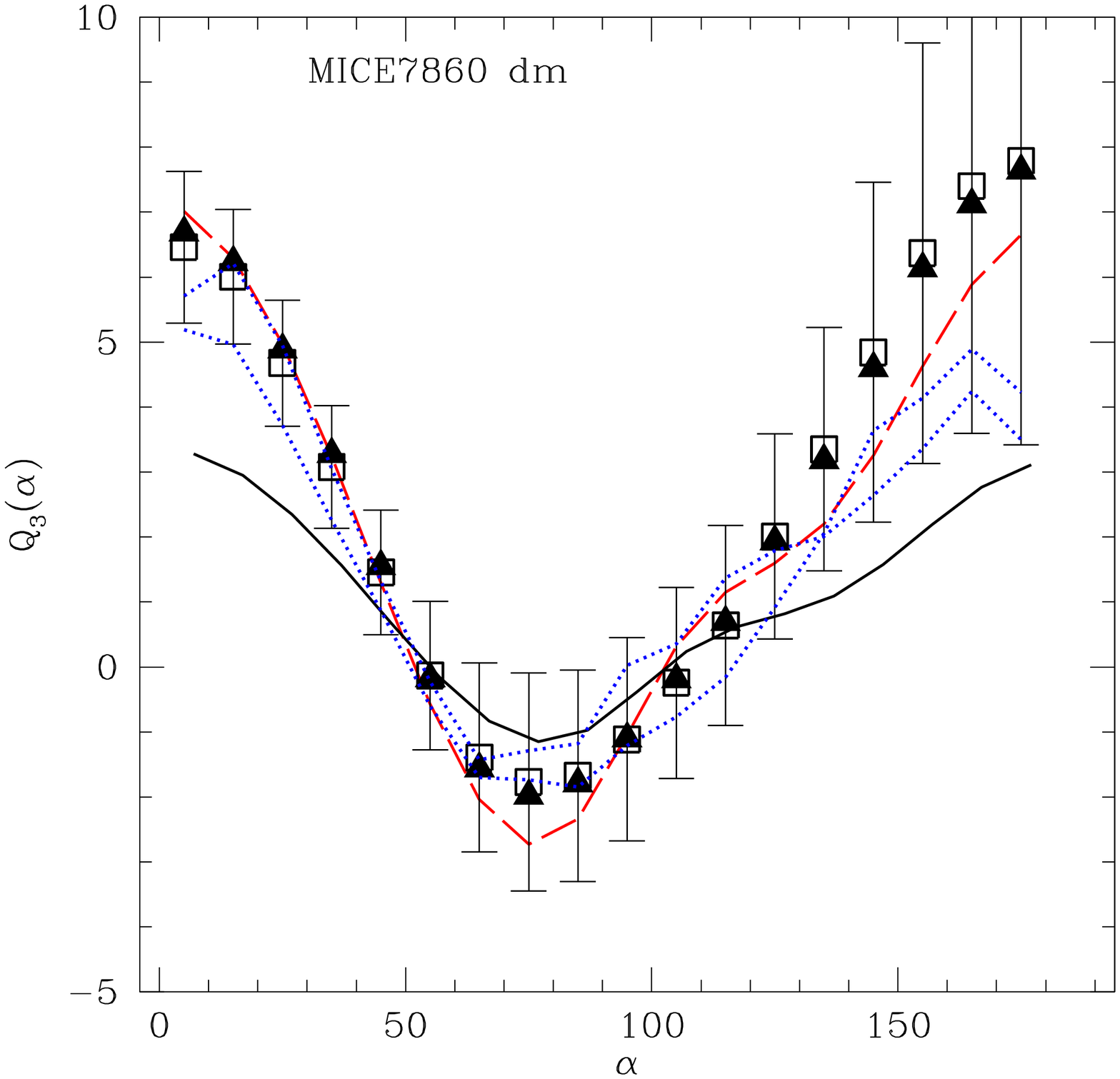}
\includegraphics[width=7.cm]{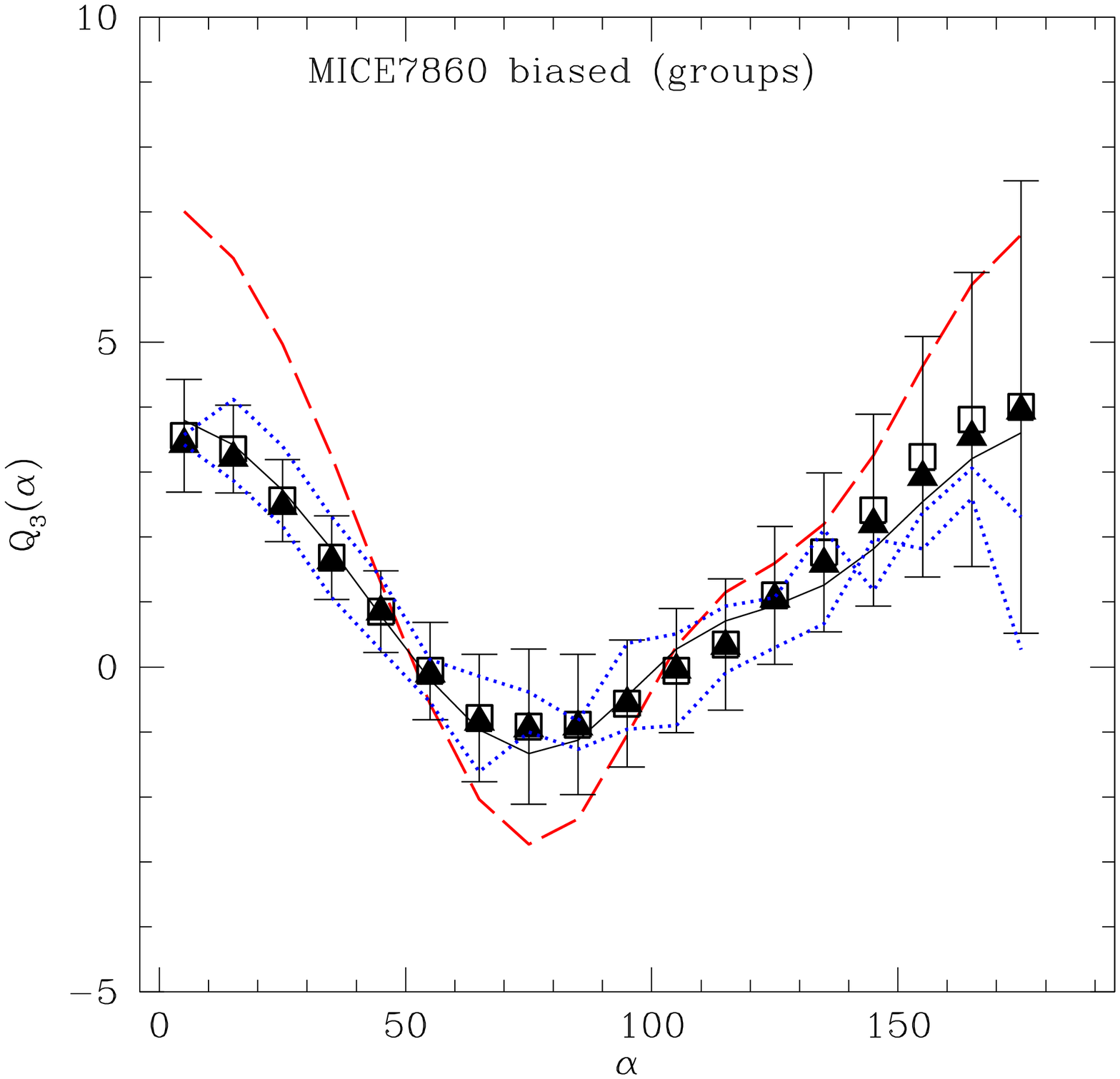}
\caption{Measurements of $Q_3$ as in Fig.\ref{fig:q3th}
in dark matter (left panel) and groups (right panel) from the MICE7680 mock
simulations. Mean and errors
are estimated from 512 subsamples of $~ 1 (Gpc/h)^3$.
This is compared to the PT predictions in real space (dashed line)
and corresponding biasing predictions for groups (continuous lines) 
Triangles and squares correspond to measurements in
real and redshift space respectively. 
Two dotted lines shows the result for two single 
$1 (Gpc/h)^3$ realizations (\#20 and \#21), in redshift space.}
\label{fig:q3sim}
\end{figure*}

\section{Simulation and errors}

To check our codes and estimate errorbars,
we have used a comoving output at z=0 of a MICE simulation, 
run in the super computer Mare Nostrum in Barcelona by MICE consortium (www.ice.cat/mice).
The  simulation contains $2048^3$ dark matter particles, 
in a cube of side 7680Mpc/h (which we call MICE7680), $\Omega_M=0.25$, 
$\Omega_b =0.044$, $\sigma_8=0.8$, $n_s=0.95$ and $h=0.7$. 
This simulation uses  the power spectrum of
 \citet{eisensteinhu} (EH from now on). To be consistent, the same
approximation has also been used for the predictions.
We assume here that this EH fit
is good enough approximation for the precision in our analysis, but we note that
\citet{baughbao} found that on the BAO scale this approximation 
is only accurate at the few percent level. We also use the EH fit with ``no wiggles''
which for each cosmological model produces a match to the smoothed shape of the power
spectrum without the BAO wiggles. This is equivalent to the removal of the BAO
peak in the correlation function but keeping the same correlation as a based line.
Note that we have chosen to displayed our results in terms of $\Omega_m$ and
$\Omega_B$ for a fixed $h=0.7$. 

The observed LRG galaxies are at a mean cosmic time of $z=0.3$ rather than
cosmic time $z=0$ 
in our simulations. In principle,  this should result in a slightly larger amplitude
for the clustering in the simulations, because it corresponds to a later cosmic time.
But the simulations have a lower normalization, $\sigma_8=0.80$, 
than the value of $\sigma_8=0.85$ inferred for the LRG galaxies after
accounting for the effect of bias (see Paper I).
This two effects partially compensate each other and the net result is that simulations
have very similar clustering amplitude to that inferred in the LRG data.

To simulate biasing we 
select groups of particles using friend-of-friends with linking scale of $0.20$.
We find a total of 107 million groups with more than 5 particles 
($M>1.87 \times 10^{13}$). These groups approximately correspond to DM halos when the number
of particles in the group  is larger than few tens of particles \cite{gus08}. 
When the number of particles is smaller than this,  the group might not 
always correspond a virialized DM halo.  But in any case, these groups sample high density
regions which are biased tracers of the dark matter distribution.
We can produce different mock galaxy catalogs by choosing the group
richness (ie  mass or number of particles). Very massive groups are more biased and
more rare than smaller groups. What we do here is to select the group
mass cut-off to reproduce the galaxy clustering in the LRG observations. It
turns out that when we do this we get a number density of groups
which is quite similar to the number density of  LRG galaxies. 
Our mock catalogs are not realistic in the sense that we
have not simulated the physics of galaxy formation. But we only use
these catalogs for error estimation. Errors depend on the statistical properties
of the simulation and not on process that produce such statistics.
For our purposes here, all we need from mock simulations is that they have
approximately the same volume, bias $b$, number density and 
similar values of $\xi_2$ and $\xi_3$ that observations. Errors
only depend on this quantities.

We have also used a MICE simulation with $2048^3$ dark matter particles, 
in a cube of side 3072Mpc/h (which we call MICE3072, same parameters as MICE7680)
which has 15 times better mass resolution to check for mass resolution effects.
We find very similar results in both cases, but obviously MICE7680 provides
more sampling volume to estimate reliable errors.

When we  select groups with $M > 2-4 \times 10^{13}$,
both the clustering amplitude ($b_1 \simeq 1.9-2.2$ for $\sigma_8=0.8$) 
and the number density ($\bar{n} \simeq 4-6 \times 10^{-5}$) 
are similar to the real LRG galaxies in our SDSS sample (the range reflects
the fact that the actual number depend on LRG sample used). As the mock simulations
are similar to the real data we will use the variance of clustering in the simulations
 to estimate errorbars for our analysis. The resulting errors from simulations are 
typically in good agreement with JK errors from the real data (see paper I for details).
We will also check
if we can recover the theory predictions for $Q_3$ by using mock simulations with similar
size as the real data.

We have divided the big MICE7680 cube in $8^3$ sub-cubes, each of side 961Mpc/h and apply
the LRG SDSS mask to obtain $N_M=512$ LRG mocks from both dark matter and groups.
From the $N_M$ mock catalogs, we can estimate what we call the 
Monte Carlo (MC) covariance  matrix:

\begin{equation}
\label{eq:covMC}
 C_{ij} =  \frac{1}{N_m} \sum_{k=1}^{N_m}(\xi(i)^{k}-\widehat\xi(i))(\xi(j)^{k}-\widehat\xi(j))
\end{equation}

where $\xi(i)^{k}$ is the measure of $\xi_2$, $\xi_3$ or $Q_3$ 
in the k-th mock simulation ($k=1,...N_m$)  and $\widehat\xi(i)$ is the mean over 
$N_m$ realizations. 
The case i=j gives the diagonal error (variance).\

\begin{figure}
\includegraphics[width=7.5cm]{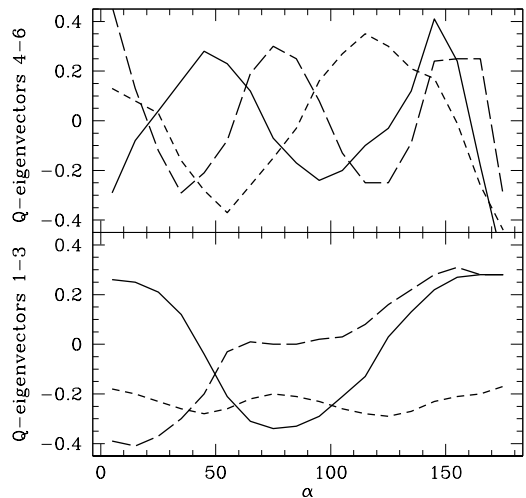}
\caption{ 
Here we show the first 6 eigenvectors in the SVD using 512 subsamples in MICEL7680.
Bottom panel shows the first 3 principal components (dashed, continuous and
long-dashed lines) and the top panel shows the next 3 components, ranked by amplitude. 
While the $\Omega_m -\Omega_B$ constraints are degenerate using the first 3 
components, this degeneracy is broken by using more eigenvectors because 
they can separate BAO features.}
\label{fig:eigenv}
\end{figure}

\begin{figure*}
\includegraphics[width=7cm]{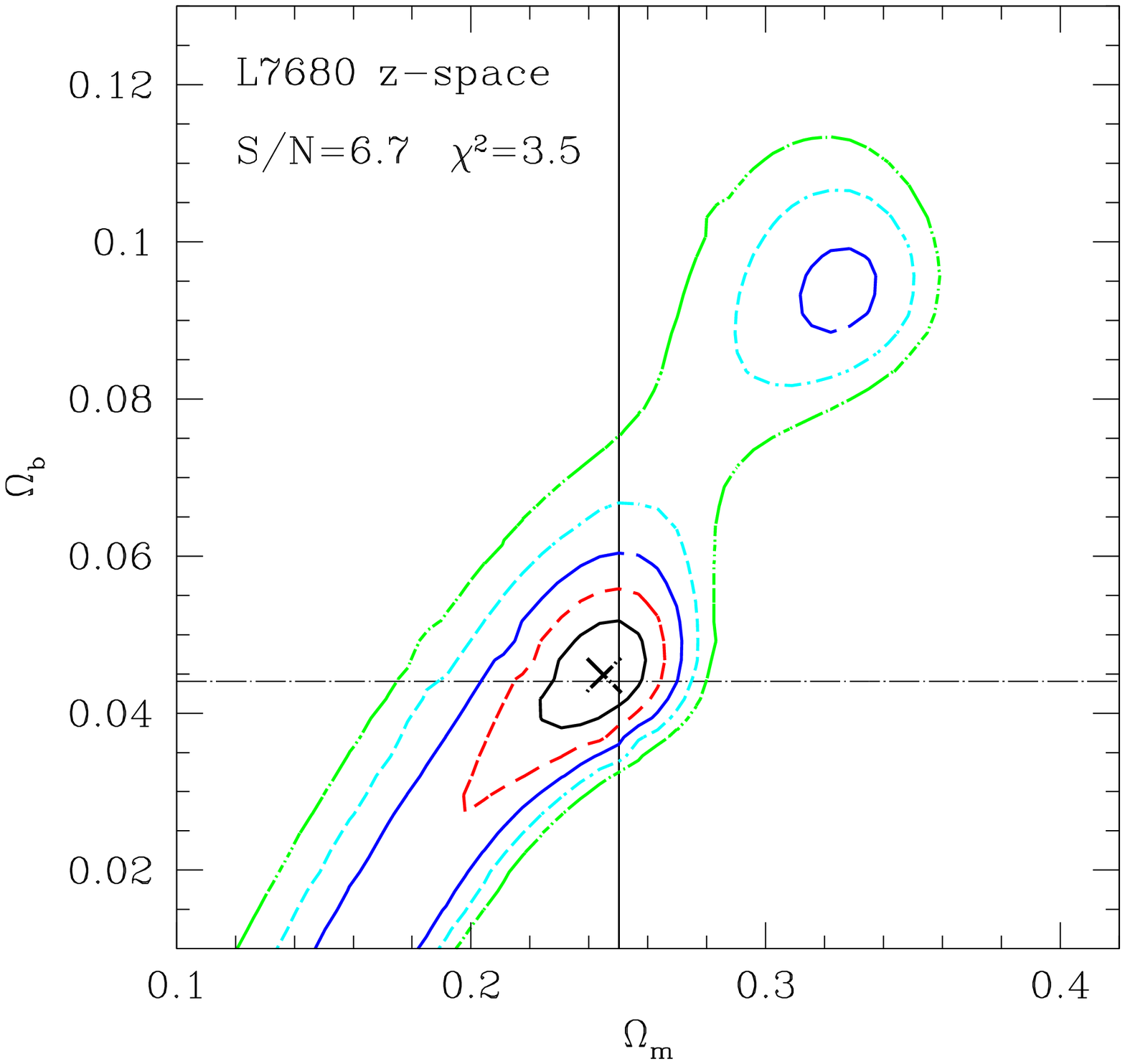}
\includegraphics[width=7cm]{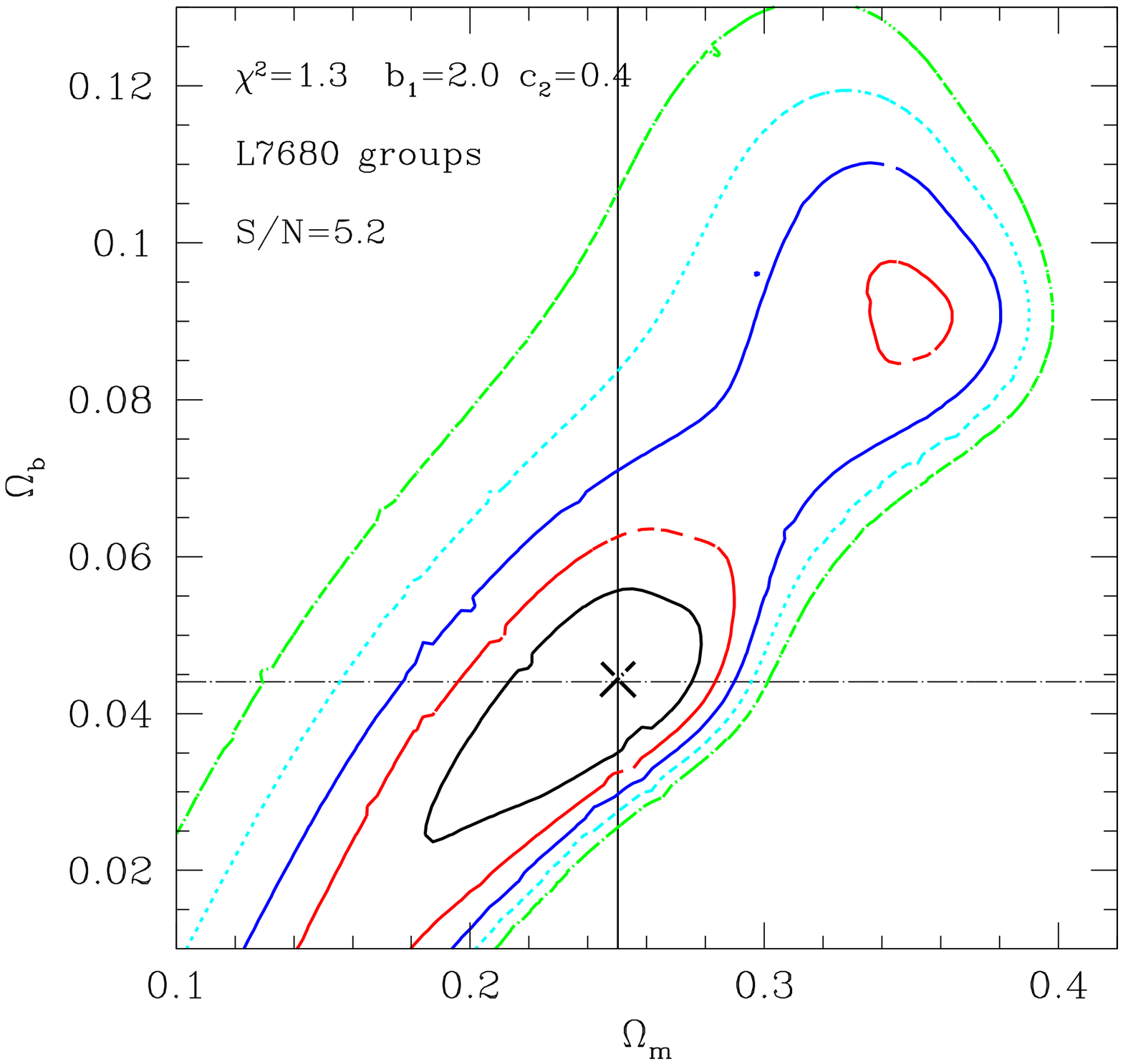}
\caption{Contours of constant $\Delta \chi^2 = 1, 2.3, 4, 6.2.$ 
and $9$ obtained from fitting models to measurements in dark matter (left panel)
and groups (right panel) in MICE7680 mock simulations 
using SVD with MC covariance matrix corresponding to
a Survey of about 1 $Gpc^3/h^3$.
The crossing lines show
the input values in the simulations, while the cross 
shows the best fit value. At 1-sigma, 
best fit is in excellent agreement with the input. But 
at 2-sigma there is a secondary peak and a strong
degeneracy in the $\Omega_B-\Omega_m$ plane.
}
\label{fig:OmObsim}
\end{figure*}

Fig.\ref{fig:q3sim} compares the mean and errors in the MICE7680 mocks
with the PT predictions (dashed lines) in real space. 
At these large scales, results in real
space are almost identical to redshift space for both
dark matter (left panel) and groups with $M>1.87 \times 10^{13}$ (right panel).
The PT predictions and the biasing predictions
work remarkably well. For bias we have used $b_1=1.9$ and $c_2=0.2$ (continuous lines).
The value $b_1=1.9$ is estimated empirically from the ratio of the 2-point function
of the groups to that of dark matter. This ratio is fairly constant for $r>11 Mpc/h$.
The non-linear bias $c_2$ produces a global shift that
we have just fitted to the data. These estimated values of $b_1$ and $c_2$
agree very well with halo model prediction \cite{bhalo} for the mass of these groups. 
We note that we show the mean of the 512 mocks with errors from the dispersion
in  redshift space, which are slightly smaller (10-30\%) 
than in real space. Dotted lines show results from two of the 512 realizations,
illustrating the strong covariance in $Q_3(\alpha)$.

We apply a $\chi^2$ method to fit the models
by inverting the covariance matrix $C_{ij}$, 
as explained in \citet{gaztascocci2005}.
Before inverting $C_{ij}$ in Eq.(8), notice that the values of $C_{ij}$  
are estimated in practice to within a limited resolution in Eq.\ref{eq:covMC}, 
$\Delta C_{ij} \simeq \sqrt{2\over{N_m}}$. Therefore if the number 
of mocks $N_m$ is small or if 
there are degeneracies in $C_{ij}$, the inversion will be affected by numerical 
instabilities. In order to eliminate this problem, we perform a Singular 
Value Decomposition (SVD) of the matrix.
 By doing the SVD decomposition, we can choose the number of modes 
 we wish to include in our $\chi^2$ by effectively setting the corresponding 
inverses of the small singular values to zero. In practice, we work only 
with the subspace of ``dominant modes"  which satisfy
$\lambda_i^2> \sqrt{2/N_m}$ which is the resolution to which we can estimate 
the covariance matrix elements. Typically results converge after using a few
 singular values. Most of the times there is no gain in using more components
 which indicates that the effective number of degrees of freedom is smaller than
 the number of bins. Sometimes
the fit gets bad when using larger number of components, which indicates
instabilities in the covariance inversion. The results presented here are
quite robust. We find that they are basically the same when we use different
covariance matrices, corresponding to samples with different amplitude
of clustering or different shot-noise, within the range that roughly matches the data.
In particular, results are quite similar when we use DM particles mocks instead 
of groups mocks to estimate $C_{ij}$. The absolute errors in $Q_3$ are roughly the
same for dark matter and groups, despite the difference in the amplitude.
This give us confidence that what we find in real data
is not an artifact of our analysis.

Fig.\ref{fig:eigenv} shows the first 6
eigenvectors or principal components, which resemble an harmonic
decomposition of $Q_3(\alpha)$, as proposed by \citet{szapudi2004}.
The first component is roughly a constant, like a monopole, the second
component corresponds to the ''V'' shape mentioned above, like a  quadrupole,
and the third (short-dashed line)  is similar to a dipole.
The next components resemble higher multipoles and can be combined
to define localized structure in $Q_3(\alpha)$. With the first two components
it is not possible to break the  $\Omega_B-\Omega_m$ degeneracy, given 
the errorbars.  We have checked this in both the LRG data and the simulations.
As we increase the number
of multipoles we can see how the degeneracy in the  $\Omega_B-\Omega_m$
plane begins to break down. The BAO feature only shows in the higher components
and breaks the $\Omega_B-\Omega_m$ degeneracy for large values of $\Omega_B>0.03$.
Lower values show no significant BAO peak, given the errors, and this results in
a strong degeneracy in  $\Omega_B-\Omega_m$ . We have played with the models and find that even if
we artificially reduce the errors we need more that 2 eigenvalues to
have a $\Omega_B$ detection.

In Fig.\ref{fig:OmObsim} we show how well we can recover the values
of $\Omega_m-\Omega_b$ from the mocks shown in Fig.\ref{fig:q3sim}. 
We use 9 singular values, as in the LRG data, but results are quite
similar from 5 to 9 singular values. 
In the case of dark matter (left panel in Fig.\ref{fig:OmObsim}), 
we do not include biasing parameters.
In the case of groups (right panel) the results are marginalized for
$b_1=1.7-2.2$ and $c_2=0.0-5.0$ . The best fit value is in excellent agreement
with the input values both for $\Omega_m-\Omega_b$ and for $b_1-c_2$.
This illustrates that fitting
$b_1$ and $c_2$ from $Q_3$ in groups could be used to roughly estimate the mass of the groups
or galaxy clusters. But even when we recover well the input value,
note how errors are large and produce degenerate
values at 2-sigma level for the size of our mock sample. As we will see below, this is 
because $\Omega_b$ is low in the simulation. The volume of data in a single mock, of
about 1 (Gpc/h)$^3$,
is not large enough to break the $\Omega_m-\Omega_b$ degeneracy for low values of $\Omega_b$. 
This could be improved by using more shape configurations in $Q_3$ or smaller smoothing scales
(we use cubic pixels with 11 Mpc/h on a side). The second (weaker) minimum $\chi^2$
that shows in Fig.\ref{fig:OmObsim} for larger values of $\Omega_b \simeq 0.09$ 
corresponds the case where the BAO peak moves to smaller values of $\alpha$ and overlaps with
the valley of the "V" shape in $Q_3(\alpha)$. The $Q_3$ amplitude is low and the relative error 
is large in this region. This allows a BAO peak to be compatible with the simulations.
This will also affect our LRG measurements in observations, although in this case the
relative error is much smaller thanks to the non-linear bias $c_2$ which increases
the overall amplitude of $Q_3(\alpha)$ and allows a more significant detection
of $\Omega_b$.

\section{Data and Analysis}

\subsection{Data Sample}

The luminous red galaxies (LRGs) are selected by color and magnitude to obtain
intrinsically red galaxies in Sloan Digital Sky Survey (SDSS). See
\citet{eisenstein2001} or http://www.sdss.org for a complete description of the
color cuts. These galaxies trace a big volume, around $1Gpc^3h^{-3}$, which make
them perfect to study large scale clustering. LRGs are red old
elliptical galaxies, which are usually passive galaxies, with relatively low
star formation rate. Since they reside in the centers of big halos
they are highly bias with $b \simeq 2$, between regular galaxies and
clusters.

LRG's are targeted in the photometric catalog, via cuts in the (g-r, r-i, r) 
color-color-magnitude cube. Note that all colors are measured using model 
magnitudes, and all quantities are corrected for Galactic extinction 
following \cite{ext1998}.
The galaxy model colors are rotated first to a basis that is aligned with 
the galaxy locus in the (g-r, r-i) plane according to:

$c_{\perp}$= (r-i) - (g-r)/4 - 0.18

$c_{||} $= 0.7(g-r) + 1.2[(r-i) - 0.18]\\

Because the 4000 Angstrom break moves from the g band to the r band 
at a redshift z $\simeq$ 0.4, two separate sets of selection 
criteria are needed to target LRGs below and above that redshift:\\

Cut I for z $<$ 0.4

    rPetro $<$ 13.1 + $c_{||}$ / 0.3

    rPetro $<$ 19.2

    $|c_{\perp}|$ $<$ 0.2

    $mu_{50}$ $<$ 24.2 mag $arcsec^{-2}$

    $r_{PSF}$ - $r_{model}$ $>$ 0.3 \\

Cut II for z $>$ 0.4

    $r_{Petro}$ $<$ 19.5

     $|c_{\perp}|$ $>$ 0.45 - (g-r)/6

    g-r $>$ 1.30 + 0.25(r-i)

    $mu_{50}$ $<$ 24.2 mag $arcsec^{-2}$

    $r_{PSF}$ - $r_{model}$ $>$ 0.5 \\

Cut I selection results in an approximately volume-limited LRG sample 
to z=0.38, with additional galaxies to z $\simeq$ 0.45. Cut II selection adds 
yet more luminous red galaxies to z $\simeq$ 0.55. The two cuts together 
result in about 12 LRG targets per $deg^2$ that are not already 
in the main galaxy sample (about 10 in Cut I, 2 in Cut II).
The radial distribution and magnitude-redshift diagrams for these
galaxies are shown in Paper I \cite{paper1}.

We k-correct the r magnitude using the Blanton program 'kcorrect'
\footnote{http://cosmo.nyu.edu/blanton/kcorrect/kcorrect\_help.html}. We need to
k-correct the magnitudes in order to obtain the absolute magnitudes and
eliminate the brightest and dimmest galaxies. We have seen that the previous
cuts limit the intrinsic luminosity to a range $-23.2<M_r<-21.2$, and we only
eliminate from the catalog some few galaxies that lay out of the limits. Once we
have eliminated these extreme galaxies, we still do not have a volume limited
sample at high redshift. For the 2-point function analysis we  account for 
this using a random catalog with identical selection function but 20 times
denser (to avoid shot-noise) . The same is
done in simulations. Computationally this is very time consuming (specially
as there are 512 simulations) because it involves $N^2$ operations
where $N \simeq 10^6$. For the 3-point, using the random
catalogs would involve $N^3$ operations which begins to be very challenging with
current computer power.  We will therefore use a different estimator based
on a pixelization of the sample. This estimator is ideally match to a volume limited
sample, where the full volume is equally sampled and there is no radial selection
function (we still have angular mask and radial boundaries).

For the 3-point function analysis presented in the paper 
we select a volume limited sample with $-22.5<M_r<-21.5$ and $z=0.15-0.38$ 
from the spectroscopic sample of LRG in the SDSS DR6.  We choose this particular
sample because it is the best compromise between volume and number density.
There are about
$40,000$ LRG galaxies in this sample ($\bar{n} \simeq 4 \times 10^{-5}$).

\subsection{Correlation functions}

We  estimate the correlation functions with a 
fast algorithm described in some detail in 
\citet{barrigagazta2002,gazta2005}.
This algorithm allows a fast calculation of two and three-point function for millions
of points. 
The first step is to discretize the simulation box
into $Lsize^3$ cubic cells. We assign each particle to a node of this new
latticed box using the nearest grid point particle assignment.  
We precalculate the list of relative neighbors to any given 
node in the lattice. To compute now the two-point and three-point correlation 
functions we use:
\bea{eq:x2x3}
\xi_2(r_{12}) &=& 
{\sum_{i,j} \delta_i \delta_j \over{ \sum_{i,j} 1}}
\\
\xi_3(r_{12},r_{23},r_{13}) &=&  {\sum_{i,j,k} \delta_i \delta_j \delta_k 
\over{ \sum_{i,j,k} 1}}
\eea
where $i$ extends over all nodes in the lattice, $j$ over the list of
precalculated neighbors that are at a distance $r_{12} \pm dr/2$ from $i$, and
$k$ is over the neighbors at distance $r_{23} \pm dr/2$ from $j$ and
$r_{13}\pm dr/2$ from $i$. We take $dr$ to be equal to
the pixel size.

There are two sources of errors in this estimation: a) {\bf Shot-noise}
which scales as one over the square root of the number of
pairs or triplets in each bin b) {\bf sampling variance} which
scales with the amplitude of the correlations.
It is easy to check that for the size and density of our
sample the shot-noise term dominates over the sampling variance
error. This has been checked in detailed by 
using DM simulations which have a large density and can be diluted 
to explore the shot-noise contribution to the error budget
(see also Paper I). 

We choose cubical pixels of $dr=11$ Mpc/h on the side. This is an
adequate compromise to measure BAO. We want this number
to be as large as possible to reduce shot-noise and to average
over many triangles in a fast way. On the other hand, we need a
good resolution to avoid loosing too much shape and BAO information.
Once the pixel size is fixed to $dr=11$ Mpc/h we are forced to use $r_{12}>3 dr$ 
which is the smaller distance that does not distort the $Q_3$ shape 
\cite{barrigagazta2002}. To reach the BAO scale we need $r_{13}>8 dr$.
So this fixes our choice of triangles.
The 2-point function for this sample and pixelization is shown 
in Fig.B11 of Paper I \cite{paper1}.

\subsection{Results}

\begin{figure}
\includegraphics[width=7.cm]{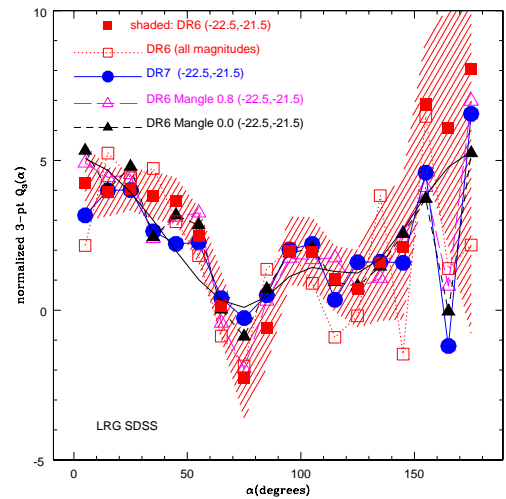}
\caption{Estimations of $Q_3$ as in Fig.\ref{fig:q3th} from observations using different
samples and masks. Closed (red) squares and (red) shaded region
show the main measurements and errors used in this paper, ie for a volume limited 
sample from DR6 with a magnitude range $-22.5<M<-21.5$ and $0.15<z<0.38$.
Closed (blue) circles show the corresponding result in the DR7 sample.
Open and closed triangles show the DR6 results using the 
 MANGLE mask of Swanson et al. (2008) with greater than 0.0 or 0.8 completeness fractions.
Open squares use all magnitudes. The (black) continuous line correspond to our best
fit model in Fig.\ref{fig:q3S21}}
\label{fig:q3sys}
\end{figure}

After the release of DR6, 
\citet{swanson08} provided mask information in a readily usable form, translating 
the original mask files extracted from the NYU Value-Added Galaxy Catalog \citep{blanton2005}, 
from MANGLE into Healpix format \citep{healpix}. \citet{paper1} describe 
how they constructed a survey "mask" for LRGs and tested the impact of the mask on 
clustering measurements using mock catalogs. Using the same techniques, we have 
also examined the correlation function of LRGs in DR7, which has become available 
since the submission of \citet{paper1}. In Fig.~\ref{fig:q3sys}
we plot a summary  of the possible systematics in the estimation of the
$Q_3$ measurements. Our main result, that will be used for comparison with models,
is shown as a shaded region, corresponding to the 1-sigma region. Results using different masks
or the DR7 release are quite similar despite variations of $\sim 17\%$ in 
the fraction of galaxies or area used in the different cuts. We also show comparison to  
the results using all galaxies (open squares), rather than just a volume limited subsample. In this
case we define density fluctuations using the local mean density provided by the random catalogs,
rather than the overall mean density as we do for volume limited catalogs. The result is noiser
(because of the shot-noise in the random catalogs) but in very good agreement with the other
measurements. Evolutionary effects (that change the mean density)
or magnitude effects do not seem important in this sample.
We conclude that the results are very robust to the systematic variations 
that we
have tried, and choose to use our default DR6 volume limited sample because this is the
one that have been more tested and is the bases for our error analysis.

\begin{figure}
\includegraphics[width=7.cm]{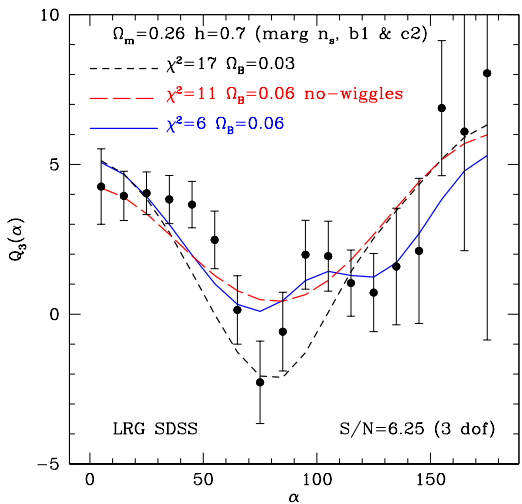}
\caption{Values of $Q_3$ as Fig.\ref{fig:q3th}, this time comparing
three of the biased models (lines) to observations in LRG galaxies (symbols with
errorbars).
The total signal-to-noise in this detection is 6.25.
Models have $\Omega_m=0.26$ and $h=0.7$.
  with  $\Omega_b=0.03$
(short-dashed lines) or $\Omega_b=0.06$ (continuous line).
Long-dashed uses EH fit with the no-wiggles
(ie no BAO peak, but otherwise equal baseline spectrum)
while the other lines includes the BAO peak in the models.
We have marginalized over biasing parameters and spectral index
and show the best fit in each case. The high $\Omega_b=0.06$
model has a minimum $\chi^2=6$ (with 3 degrees of freedom)
which is significantly smaller than the mini-mun $\chi^2=17$
for the $\Omega_b=0.03$ model.
The BAO peak shows at $\alpha=100$ in both the model with $\Omega_b=0.06$
 and the data. The best fit model without BAO peak has  $\chi^2>11$, ie
a probability smaller than 1\% of being correct.}
\label{fig:q3S21}
\end{figure}

Fig.\ref{fig:q3S21} 
shows a comparison of $Q_3(\alpha)$ measurements in the LRG
sample with models. There is a good resemblance with what is expected from
theory, ie compared to Fig.\ref{fig:q3th}, with a peak at $\alpha \simeq 100$
deg. which resembles much the BAO peak predicted by models. As our signal
is shot-noise dominated one may wonder if this peak could be produce
by noise fluctuations.

\begin{figure}
\includegraphics[width=8.5cm]{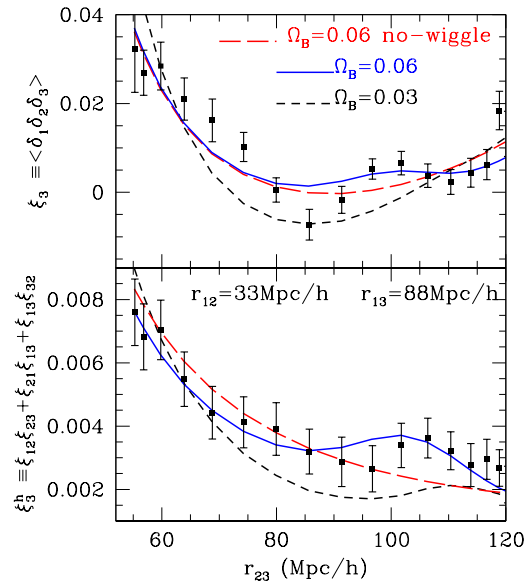}
\caption{Separate measurements of $\xi_3$ (top panel) and
hierarchical $\xi_3 \equiv \xi_2(r_{12})\xi_2(r_{23})+\xi_2(r_{21})\xi_2(r_{13})+
+\xi_2(r_{13})\xi_2(r_{32})$ (bottom panel). The models are as in Fig.\ref{fig:q3S21},
ie $\Omega_b=0.03$ (short-dashed line)  and $\Omega_b=0.06$ with 
(continuous line) and without wiggles (short-dashed lines), all with
$\Omega_m=0.26$. In this case the prediction depends not only on the biasing
parameters, but also on the $\sigma_8$ normalization. As can be seen in this
figure, the model with large $\Omega_B$ show a different shape and 
a BAO feature both in $\xi_3$ and $\xi_2$. Data follows the BAO predictions
in both quantities, as well as in $Q_3$ which is quite reassuring.}
\label{fig:x3h}
\end{figure}

An indication that this signal is real comes from Fig.\ref{fig:x3h}.
Detection of the BAO peak using the hierarchical product $\xi_3^H$ of 2-point
function, as shown in Fig.\ref{fig:x3h}, is in excellent agreement with
previous detections
\cite{detection} and results over the very same sample
in Paper I of this series \cite{paper1}. The peak seems to be detected
in all $\xi_3$, $Q_3$ and $\xi_2$ and products.

\begin{figure*}
\includegraphics[width=7cm]{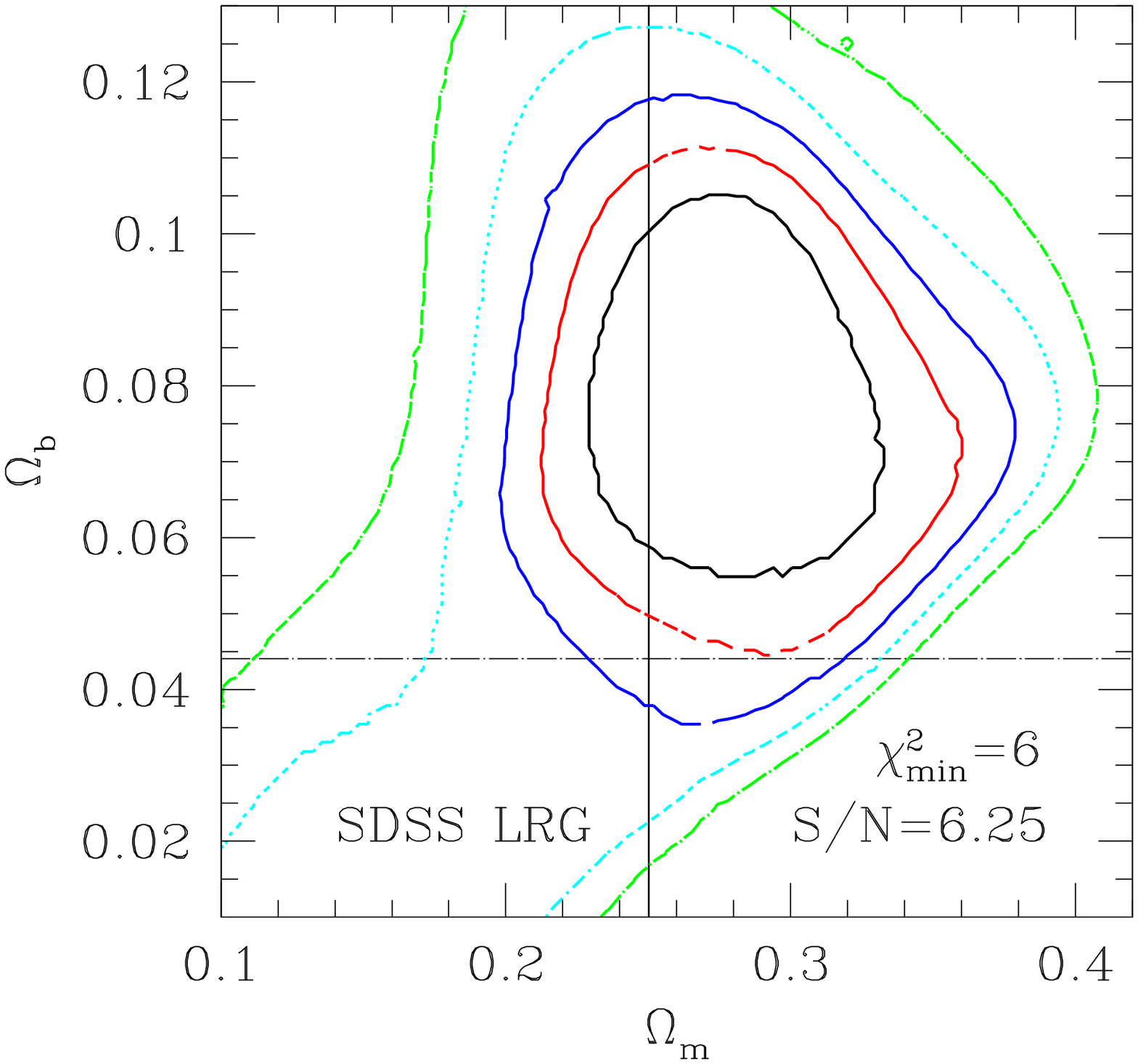}
\includegraphics[width=7cm]{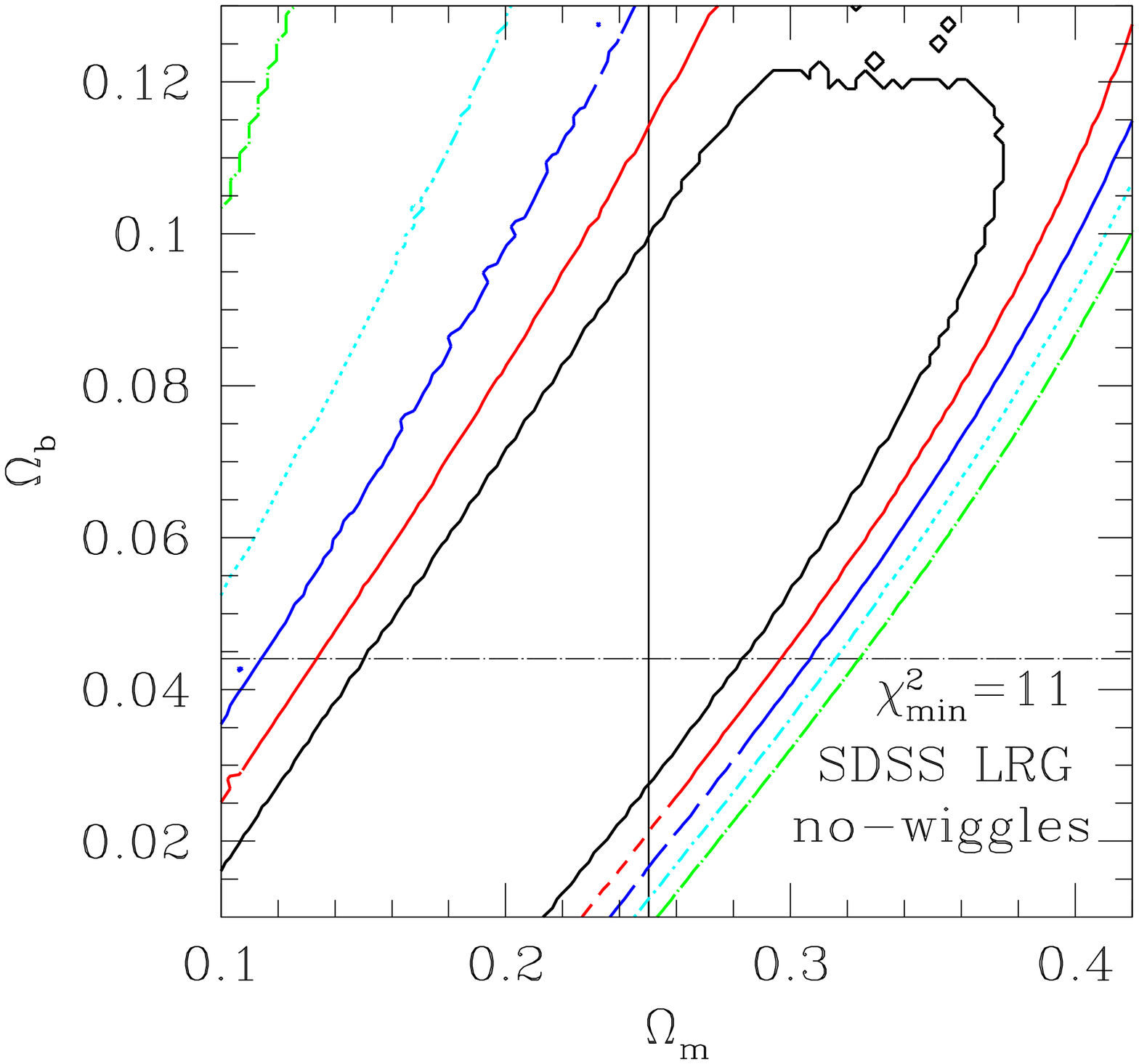}
\caption{Contours of constant $\Delta \chi^2 = 1, 2.3, 4, 6.2.$ 
and $9$ obtained from
fitting models to  data in  Fig.\ref{fig:q3S21} using a SVD
with covariance matrix from the MICE group simulations.
Contours are marginalized over $b_1=1.7-2.2$, $n_s=0.8-1.2$
and $c_2=0.0-5.0$ (we use $h=0.7$).  The crossing lines show
the best WMAP5 fit. Right panels uses EH fit with the no-wiggles
(ie no BAO peak, but otherwise equal shape)
while left panel includes the BAO peak in the models.
Best BAO fit is $\Omega_m=0.28 \pm 0.05$ and
$\Omega_B=0.079 \pm 0.025$ with $\chi^2=6$ for 3 degrees
of freedom (9 singular values minus 5 parameters in the fit).
Probability for models with no BAO peak is less that $1\%$
($\chi^2>11$).}
\label{fig:OmOb}
\end{figure*}

Left panel of Fig.\ref{fig:OmOb} shows the $\chi^2$ fit for $\Omega_B-\Omega_m$
plane using 9 singular values with a total signal-to-noise
of $6.25$. We fix $h=0.7$ and marginalized over 
spectral index $n_s=0.8-1.2$ and biasing parameters
$b_1=1.7-2.2$ and $c_2=0.0-5.0$. In the right panel we show
the same fit for the EH models with no-wiggles (ie no BAO peak).
Note how the values of $\Omega_B-\Omega_m$ become degenerate, as
expected from our previous argument that the BAO peak helps to break
the  $\Omega_B-\Omega_m$ degeneracy. The best fit value without BAO peak
is $\chi^2=11$ as opposed to $\chi^2=6$ with the BAO peak. Thus,
in relative terms the models with and without a BAO peak are between 
2 and 3-sigma away. But note that in absolute terms, models without
the BAO peak are ruled out with $>99\%$ confidence.

The WMAP5 best fit value (marked by a cross) is
outside the (2D) 1-sigma join region, but inside the (2D) 2-sigma contours.
The best fit value is for $\Omega_B=0.079 \pm 0.025$ and we find
that $\Omega_B>0.035$ at 2-sigma level for any value of $\Omega_m$.

If we fix the   $\Omega_B-\Omega_m$ to its best fit value, 
we find  $b_1 =1.7-2.2$ and $c_2=0.75-3.55$. The value of the linear
bias $b_1$ is in excellent
agreement with what we found in Paper I by fitting redshift
space distortions in the 2-point function, but the error here
is larger. The value of the
non-linear bias $c_2 \simeq 2$ is higher than the one we found in previous
section for halos $c_2^h \simeq 0.2$. This is not surprising as it is well known that more
than one LRG can occupy a single halo, in which case 
$c_2$ tends to be larger for a given $b_1$ \cite{bhalo}.
Also note that a larger value of $c_2$ makes the $Q_3$ signal-to-noise
larger in the LRG data than in the MICE7680 group mocks. This helps
defeating the shot-noise and improves
the significance of the BAO detection.

\section{Conclusions}

We have studied the large scale 3-point correlation function for luminous red
galaxies from SDSS, and particularly the reduced $Q_3=\xi_3/\xi_2^2$, which
measures the scaling expected from non-linear couplings. We find a well-detected
peak at 105Mpc/h separation that is in agreement with the predicted position of
BAO peak. This detection is significant since it is also imprinted in $\xi_2$
and $\xi_3$ separately. We focus our interpretation in $Q_3$ because
it is a measure independent of time, $\sigma_8$ or growth factor. It only depends
on the shape of the initial 2-point function and the non-linear coupling 
of the gravitational interaction. Our result for $Q_3$ is in excellent 
agreement with predictions from Gaussian initial conditions.
When we use the $Q_3$ data alone (with no fit
to the 2-point function) we are able to break the strong degeneracy between
$\Omega_m$ and $\Omega_B$ (see Fig.\ref{fig:OmOb}). 
Our detection shows a clear preference for a
high value of $\Omega_B=0.079 \pm 0.025$. This value is larger, but still
consistent at 2-$\sigma$ with recent results of WMAP ($\Omega_B=0.045$).
At 3-sigma level, the $\Omega_m-\Omega_B$ becomes degenerate.
Models with no BAO peak are ruled out at 99\% confidence level.

We have used very large realistic mock simulations to study the errors. These
simulations show  that $Q_3$ is not significantly modified in redshift
space so we can use real-space perturbation theory (see Fig.\ref{fig:q3sim}).
This agreement also indicates that loop corrections are small on BAO
scales \cite{Bernardeau08}.
This analysis is independent from 2-point statistics, which tests the linear
growth of gravity, since 3-point statistics test the non-linear growth.
A high value for $\Omega_B$ is also consistent with the
analysis of the peak in the 2-point correlation function shown in \citet{detection}
and in Paper IV (\citet{paper4}) of this series,
which detect a slightly higher peak than expected. We have done
all the analysis with just one set of triangle configurations, with fixed sides of
$r_{12}=33 \pm 5.5$ Mpc/h and $r_{13}=88 \pm 5.5$ Mpc/h, 
to center our attention to BAO scale. This is about optimal, but we notice that
there is much more to learn from $Q_3$, which will be presented in future
analysis.  Results on smaller scales are consistent with what we find here.

Data is in excellent
agreement with Gaussian initial conditions, for which $Q_3=0$.
But note that our quadratic bias detection  $c_2=0.75-3.55$
is degenerate with a primordial non-Gaussian (hierarchical) contribution.
Indeed the mean value of $c_2 \simeq 2$ seems larger in observations than in
halo simulations, for which we find $c_2^h \simeq 0.2$. 
We believe that this indicates that halos
are sometimes occupied by more than one galaxy, which increases
the effective value of $c_2$ \cite{bhalo}. But if we 
are conservative we can not rule out a primordial
non-Gaussian contribution
in the range $Q_3(Primordial)= c_2-c_2^h=[0.55,3.35]$.

Note that we have pixelized our data in cubical cells of side $dr=11$ Mpc/h.
This results in some lost of small scale information but allows for
a very fast method to estimate 3-point function \cite{barrigagazta2002}.
This is important for data, but more for simulations.
In the MICE7680 simulation there are close to $N=10^{11}$ particles. A brute
force method to estimate 3-point correlation would require $N^3=10^{33}$ 
operations, while our method based on pixels just needed $5 \times 10^{12}$
operations. 

\begin{figure}[t]
\includegraphics[width=7cm]{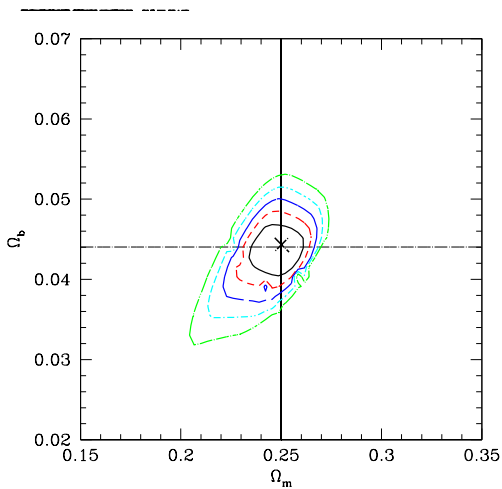}
\caption{Same as Fig.\ref{fig:OmObsim}
 for a future photometric Survey with photo-z error of $\Delta z<0.003 (1+z)$,
volume of $V = 10 Gpc^3/h^3$ and number density of $\bar{n}=10^{-3} h^3/Mpc^3$ LRG galaxies.}
\label{fig:OmObPAU}
\end{figure}

Future surveys will be able to improve much upon our measurement
here. A photometric survey with $\Delta z<0.003 (1+z)$ precision (corresponding
to $dr<9$ Mpc/h at z=0), such as in the PAU Survey \cite{PAU} should have enough 
spatial resolution to measure $Q_3(\alpha)$ as presented in this paper (recall that
we are binning our radial distances in $dr=11$ Mpc/h). Such
survey could sample over 10 times the SDSS DR6 volume (ie to z=0.9) with 20 times better LRG
number density (ie for $L>L_*$).
Fig.\ref{fig:OmObPAU} shows the forecast for such a survey, which we have
simulated with the MICE7680 mocks in redshift space with a photo-z
of $\Delta z<0.003 (1+z)$ and for the same triangles as shown in our
SDSS analysis. This is just illustrative, as we have not marginalized over
biasing and other cosmological uncertainties. But note  that 
the improvement is substantial and shows the potentiality
of this method to constrain cosmological parameters and models of structure
formation.

%\section*{Acknowledgments}

EG wish to thank Bob Nichol for suggesting the test with the EH no-wiggle model.
We acknowledge the use of simulations from the MICE consortium 
(www.ice.cat/mice) developed at the MareNostrum supercomputer
(www.bsc.es) and with support form PIC (www.pic.es),
the Spanish Ministerio de Ciencia
y Tecnologia (MEC), project AYA2006-06341 with
EC-FEDER funding, Consolider-Ingenio PAU project CSD2007-00060
and research project 2005SGR00728
from Generalitat de Catalunya. AC acknowledge support
from the DURSI department of the Generalitat de
Catalunya and the European Social Fund.

%\bibliography{tesi}

\end{document}